\documentclass[preprint]{elsarticle}
\usepackage{amssymb}
\usepackage{amsmath}
\usepackage{amsfonts}
\usepackage{ulem}
\usepackage{rotating}
\usepackage[font=small,format=plain,labelfont=bf,up,textfont=it,up]{caption}

\begin{document}
	\title{{\bf Solutions of the three-dimensional radial Dirac equation from the Schr\"odinger equation with one-dimensional Morse potential} \footnote{\bf To appear in Physics Letters A (2017) - 10.1016/j.physleta.2017.04.037}}
	
	\author[1]{M. G. Garcia}
	\ead{marcelogarcia82@gmail.com}
	\author[2]{A. S. de Castro}
	\ead{castro@pq.cnpq.br}
	\author[3]{P. Alberto}
	\ead{pedro.alberto@uc.pt}
	\author[4]{L. B. Castro}
	\ead{luis.castro@pq.cnpq.br}
	
	\address[1]{UNICAMP, Universidade Estadual de Campinas, Departamento de Matem\'{a}tica Aplicada, IMECC, 13081-970, Campinas, SP, Brazil}
	\address[2]{UNESP, Universidade Estadual Paulista, Campus de Guaratinguet\'{a}, Departamento de F\'{\i}sica e
		Qu\'{\i}mica, 12516-410, Guaratinguet\'{a}, SP,  Brazil}
	\address[3]{CFisUC, University of Coimbra, Physics Department,
		P-3004-516, Coimbra, Portugal}
	\address[4]{UFMA, Universidade Federal do Maranh\~{a}o,
		Campus Universit\'{a}rio do Bacanga, Departamento de F\'{\i}sica, 65080-805, S\~{a}o Lu\'{\i}s, MA,
		Brazil}

	\begin{abstract}
		New exact analytical bound-state solutions of the radial Dirac equation in 3+1 dimensions for
		two sets of couplings and radial potential functions are obtained via mapping
		onto the nonrelativistic bound-state solutions of the one-dimensional
		generalized Morse potential. The eigenfunctions are expressed in terms of
		generalized Laguerre polynomials, and the eigenenergies are expressed in
		terms of solutions of equations that can be transformed into polynomial
		equations. Several analytical results found in the literature, including the
		Dirac oscillator, are obtained as particular cases of this unified approach.
	\end{abstract}
	
	\begin{keyword}Dirac equation\sep%
		Morse potential%
		
	\end{keyword}
	
	\maketitle
	\section{Introduction}

In nonrelativistic quantum mechanics there are several potentials with
physical interest that allow for exact solutions, thus offering the
possibility of extracting physical information in a way which is not
possible otherwise. Among them is the generalized Morse potential $%
Ae^{-\alpha x}+Be^{-2\alpha x}$ \cite{b2tez}-\cite{b4ard1}, the singular
harmonic oscillator (SHO) $Ax^{2}+Bx^{-2}$ \cite{b4bag}, \cite{b1lan}-\cite%
{b5asc}, and the singular Coulomb potential (SCP) \thinspace $%
Ax^{-1}+Bx^{-2} $ \cite{b4bag}, \cite{b1lan}-\cite{b1haa}, \cite{b5don1},
\cite{b5ikh}, \cite{con}-\cite{das}, which have played an important role in
atomic, molecular and solid-state physics.

In a recent paper \cite{new}, it was shown that nonrelativistic bound-state
solutions of the well-known SHO and SCP in arbitrary dimensions can be
systematically generated from the nonrelativistic bound states of the
one-dimensional generalized Morse potential. The method amounts to a mapping
via a Langer transformation \cite{lang}. Later, in \cite{aop} the method was
extended to a modified D-dimensional Klein-Gordon equation featuring a
vector interaction nonminimally coupled. That extension of the method used
in \cite{new} provided a unified treatment of many known relativistic
problems via a mapping onto a unique well-known one-dimensional
nonrelativistic problem, allowing to obtain exact analytical bound-state
solutions for a large class of problems such a vector-scalar SHO plus
nonminimal vector Cornell $Ax+Bx^{-1}$ potentials and nonminimal vector
Coulomb (space component) and harmonic oscillator (time component)
potentials, vector-scalar Coulomb plus nonminimal vector Cornell potentials
and nonminimal vector shifted Coulomb potentials, vector-scalar SCP plus
nonminimal vector Coulomb potentials, and also the curious case of a pure
nonminimal vector constant potential.

In the present paper, the mapping onto the nonrelativistic bound states of
the one-dimensional generalized Morse potential via a Langer transformation
is extended to the Dirac equation in 3+1 dimensions with scalar, vector and
tensor radial potentials. This extension allows to obtain exact analytical
bound-state solutions for vector-scalar SHO plus tensor Cornell potentials
and vector-scalar SCP plus tensor shifted Coulomb potentials. In many cases
these represent new solutions, not found before. In all those circumstances
the eigenfunctions are expressed in terms of the generalized Laguerre
polynomials and the eigenenergies are expressed in terms of irrational
equations, which can be cast into polynomial equations. Furthermore, a
plethora of results found in the literature obtained through a large variety
of methods can now be seen as particular cases of the present method, which
is much more straightforward.

The paper is organized as follows. In Sec. 2 we review, as a background, the
generalized Morse potential in the Schr\"{o}dinger equation. The Dirac
equation with vector, scalar and tensor couplings and its connection with
the generalized Morse potential and the proper form for the potential
functions, are presented in Sec. 3 and two different classes of bound
solutions are discussed. The isolated solutions out of the Sturm-Liouville
problem are also discussed in this section. In Sec. 4 we draw some
conclusions.

\section{Nonrelativistic bound states in a one-dimensional generalized Morse
potential}

The time-independent Schr\"{o}dinger equation is an eigenvalue equation for
the characteristic pair $(E,\psi )$ with $E\in
\mathbb{R}
$. For a particle of mass $M$ embedded in the generalized Morse potential it
reads
\begin{equation}
\frac{d^{2}\psi \left( x\right) }{dx^{2}}+\frac{2M}{\hslash ^{2}}\left(
E-V_{1}e^{-\alpha x}-V_{2}e^{-2\alpha x}\right) \psi \left( x\right) =0,
\label{sch}
\end{equation}%
where $\alpha >0$. Bound-state solutions demand $\int_{-\infty }^{+\infty
}dx\,|\psi |^{2}=1$ and occur only when the generalized Morse potential has
a well structure ($V_{1}<0$ and $V_{2}>0$). The eigenenergies are given by
(see, e.g., \cite{new}, \cite{jmc})
\begin{equation}
E_{n}=-\frac{V_{1}^{2}}{4V_{2}}\left[ 1-\frac{\hslash \alpha \sqrt{2MV_{2}}}{%
M|V_{1}|}\left( n+\frac{1}{2}\right) \right] ^{2}.  \label{ENE}
\end{equation}%
with%
\begin{equation}
n=0,1,2,\ldots <\frac{M|V_{1}|}{\hslash \alpha \sqrt{2MV_{2}}}-\frac{1}{2}.
\label{cond}
\end{equation}%
This restriction on $n$ limits the number of allowed states and requires
\hfill\break$M|V_{1}|/\left( \hslash \alpha \sqrt{2MV_{2}}\right) >1/2$ to
make the existence of a bound state possible. On the other hand, on making
the substitutions%
\begin{equation}
\hslash \alpha s_{n}=\sqrt{-2ME_{n}},\quad \hslash \alpha \xi =2\sqrt{2MV_{2}%
}\,e^{-\alpha x},  \label{xi}
\end{equation}%
the eigenfunctions are expressed as%
\begin{equation}
\psi _{n}\left( \xi \right) =N_{n}\,\xi ^{s_{n}}e^{-\xi /2}L_{n}^{\left(
	2s_{n}\right) }\left( \xi \right) ,  \label{psi}
\end{equation}%
where $N_{n}$ are arbitrary constants, and%
\begin{equation}
L_{n}^{\left( b\right) }\left( x\right) =\sum\limits_{j=0}^{n}\frac{\Gamma
	\left( n+b+1\right) }{\Gamma \left( j+b+1\right) }\frac{\left( -x\right) ^{j}%
}{j!\left( n-j\right) !},\quad b>-1
\end{equation}%
are the generalized Laguerre polynimials (see, e.g., \cite{leb}, \cite%
{abramowitz}).

\section{The Dirac equation}

The time-independent Dirac equation for a spin $1/2$ fermion with energy $%
\varepsilon $ and with mass $m$, in the presence of a potential reads (with $%
\hslash =c=1$)
\begin{equation}
\left( \vec{\alpha}\cdot \overrightarrow{p}+\beta m+\mathcal{V}\right) \Psi
=\varepsilon \Psi ,  \label{p1}
\end{equation}%
where $\overrightarrow{p}$ is the momentum operator and $\vec{\alpha}$ and $%
\beta $ are $4\times 4$ matrices which, in the usual representation, take
the form%
\begin{equation}
\vec{\alpha}=\left(
\begin{array}{cc}
0 & \overrightarrow{\sigma } \\
\overrightarrow{\sigma } & 0%
\end{array}%
\right) ,\quad \beta =\left(
\begin{array}{cc}
I_{2} & 0 \\
0 & -I_{2}%
\end{array}%
\right) .  \label{p3}
\end{equation}%
Here $\overrightarrow{\sigma }$ is a three-vector whose components are the
Pauli matrices, and $I_{N}$ stands for the $N\times N$ identity matrix. In
the following, we consider%
\begin{equation}
\mathcal{V}\left( r\right) =V_{v}\left( r\right) +\beta V_{s}\left( r\right)
+i\beta \vec{\alpha}\cdot \hat{r}U\left( r\right) .  \label{p4}
\end{equation}%
In the last term, $\hat{r}=\vec{r}/r$, and the radial functions in Eq. (\ref%
{p4}) are named after the properties their respective terms have under
Lorentz transformations: $V_{v}$ corresponds to the time component of a
vector potential, $V_{s}$ is a scalar potential, and $U$ is a tensor
potential \cite{prc_tensor}. In spherical coordinates, $\Psi $ is expressed
in terms of spinor spherical harmonics%
\begin{equation}
\Psi \left( \vec{r}\right) =\left(
\begin{array}{c}
i\dfrac{g_{\kappa }\left( r\right) }{r}\mathcal{Y}_{\kappa m_{j}}\left( \hat{%
r}\right) \\
\\
-\dfrac{f_{{\kappa}}\left( r\right) }{r}\mathcal{Y}_{\tilde{\kappa}%
m_{j}}\left( \hat{r}\right)%
\end{array}%
\right) ,
\end{equation}%
where $\kappa =\pm \left( j+1/2\right) =-\tilde{\kappa}$ are eigenvalues of
the spin-orbit operator $K=-\beta \left( 2\vec{S}\cdot \vec{L}+I_{4}\right) $%
, $j$ is the total angular momentum quantum number ($m_{j}$ refers to its
third component), and $\vec{S}$ and $\vec{L}$ are the spin and angular
momentum operators, respectively. More explicitly, the spin-orbit coupling
quantum number $\kappa $ is related to the upper component orbital angular
momentum quantum number $l$ by%
\begin{equation}
\kappa =\left\{
\begin{array}{c}
-\left( l+1\right) =-\left( j+1/2\right) ,\quad j=l+1/2\;(\kappa <0) \\
\\
l=+\left( j+1/2\right) ,\quad j=l-1/2\;(\kappa >0).%
\end{array}%
\right.  \label{p6}
\end{equation}%
The upper and lower radial functions obey the coupled first-order equations:

\begin{eqnarray}
\left[ \frac{d}{dr}+\frac{\kappa }{r}+U\left( r\right) \right] g_{\kappa
}\left( r\right) &=&\left[ m+\varepsilon -V_{\Delta }\left( r\right) \right]
f_{{\kappa}}\left( r\right)  \notag \\
&&  \label{P5} \\
\left[ \frac{d}{dr}-\frac{\kappa }{r}-U\left( r\right) \right] f_{{\kappa}%
}\left( r\right) &=&\left[ m-\varepsilon +V_{\Sigma }\left( r\right) \right]
g_{\kappa }\left( r\right) ,  \notag
\end{eqnarray}%
where we have introduced the \textquotedblleft sum\textquotedblright\ and
the \textquotedblleft difference\textquotedblright\ potentials defined by $%
V_{\Sigma }=V_{v}+V_{s}$ and $V_{\Delta }=V_{v}-V_{s}$.

It is instructive to note that the charge-conjugation operation is
accomplished by the changes of sign of $\varepsilon $, $V_{v}$, $U$ and $%
\kappa $. In turn, this means that $V_{\Sigma }$ turns into $-V_{\Delta }$, $%
V_{\Delta }$ into $-V_{\Sigma }$, $g_{\kappa }$ into $f_{\kappa }$ and $%
f_{\kappa }$ into $g_{\kappa }$. Therefore, to be invariant under charge
conjugation, the Dirac equation must contain only a scalar potential.
Furthermore, $g_{\kappa }$ and $f_{\kappa }$ should be square-integrable
functions for bound states.

Due to charge conjugation, solutions for $V_{\Sigma }=0$ can be conveniently
obtained from those ones for $V_{\Delta }=0$, provided those solutions are
analytical. These correspond, respectively, to so-called pseudospin and spin
symmetry conditions of the Dirac equation (see \cite{Liang_Meng_Zhou_rev}
for a recent review). Therefore, we concentrate our attention to the case $%
V_{\Delta }=0$. In this case, one obtains a second-order differential
equation for $g_{\kappa }$ when $\varepsilon \neq -m$ and a first-order
differential equation for $g_{\kappa }$ when $\varepsilon =-m$.

\subsection{The Sturm-Liouville problem for $V_{\Delta }=0$ ($\protect%
\varepsilon \neq -m$)}

For $V_{\Delta }=0$ and $\varepsilon \neq -m$,

\begin{eqnarray}
f_{{\kappa}}\left( r\right)&=&\frac{1}{\varepsilon+m} \left[ \frac{d}{dr}+%
\frac{\kappa }{r}+U\left( r\right) \right] g_{\kappa}\left( r\right)  \notag
\\
\frac{d^{2}g_{\kappa }\left( r\right) }{dr^{2}}&+&2M\left[ \widetilde{%
\varepsilon }-V\left( r\right) -\frac{\kappa \left( \kappa +1\right) }{%
2Mr^{2}}\right] g_{\kappa }\left( r\right) =0,  \label{Eq}
\end{eqnarray}%
The
effective energy $\widetilde{\varepsilon }$ and the effective potential $V$
are expressed by
\begin{eqnarray}
2M\widetilde{\varepsilon } &=&\varepsilon ^{2}-m^{2}  \notag \\
&&  \label{treze} \\
2MV(r) &=&(\varepsilon +m)V_{\Sigma }(r)-\frac{dU(r)}{dr}+2\kappa \frac{U(r)%
}{r}+[U(r)]^2,  \notag
\end{eqnarray}%
and $g_{\kappa }\rightarrow 0$ as $r\rightarrow \infty $ for bound-state
solutions. The positive parameter $M$ has dimension of mass and no effect on
$\varepsilon $, $V$, $f_{{\kappa }}$ and $g_{{\kappa }}$, and its presence
is justified for comparison with eq. (1).

Following Ref. \cite{new}, with effective potentials expressed by%
\begin{equation}
V\left( r\right) =Ar^{\delta }+\frac{B}{r^{2}}+C,  \label{vef}
\end{equation}%
the Langer transformation \cite{lang}
\begin{equation}
g_{\kappa }(r)=\sqrt{r/r_{0}}\,\phi _{\kappa }(x)\ ,\quad
r/r_{0}=e^{-\Lambda \alpha x}\ ,  \label{Langer}
\end{equation}%
with $r_{0}>0$, $\Lambda >0$ and $\alpha $ being as in eq. (\ref{sch}),
transmutes the radial equation (\ref{Eq}) into%
\begin{equation}
\frac{d^{2}\phi _{\kappa }\left( x\right) }{dx^{2}}+2M\left\{ -\frac{\left(
\Lambda \alpha S\right) ^{2}}{2M}-\left( \Lambda \alpha r_{0}\right) ^{2}%
\left[ Ar_{0}^{\delta }e^{-\Lambda \alpha \left( \delta +2\right) x}+\left(
C-\widetilde{\varepsilon }\right) e^{-2\Lambda \alpha x}\right] \right\}
\phi _{\kappa }\left( x\right) =0,  \label{sch2}
\end{equation}%
with%
\begin{equation}
S=\sqrt{\left( \kappa +1/2\right) ^{2}+2MB}.  \label{etil}
\end{equation}%
Comparison of eqs. (\ref{sch2}) and (\ref{sch}) shows that for $\delta =0$
or $\delta =-2$, i.e., a pure inversely quadratic potential, bound solutions
are not allowed. A connection with the bound states of the generalized Morse
potential of eq.~(\ref{sch}) is obtained only if the pair $(\delta ,\Lambda
) $ is equal either to $(2,1/2)$ or $(-1,1)$, and, as an immediate
consequence of the reality of $S$, i.e., $S^{2}>0$, one must have
\begin{equation}
2MB>-(\kappa +1/2)^{2}.  \label{beta}
\end{equation}%
Actually, if $2MB>-1/4$, the above condition will be satisfied for all
values of $\kappa $, so that the term inversely quadratic in (\ref{vef})
cannot be strongly attractive. Furthermore, since the asymptotic behaviour
of (\ref{sch2}) implies that $\phi _{\kappa }\left( x\right) \underset{%
x\rightarrow +\infty }{\rightarrow }e^{-\Lambda \alpha Sx}$ and therefore,
from (\ref{Langer}), one has
\begin{equation}
g_{\kappa }\left( r\right) \underset{r\rightarrow 0}{\rightarrow }r^{1/2+S}.
\label{ur}
\end{equation}

At this point it is already worthy remarking that the fact that i) there are
no bound solutions for a pure effective inversely quadratic potential; ii)
the determination of the critical strength of the term containing the
effective inversely quadratic potential; iii) the boundary condition $%
g_{\kappa }\left( 0\right)=0$, all emerge naturally as a consequence of the
mapping onto the one-dimensional Morse problem.

Effective potentials with the general form (\ref{vef}) are achieved by
choosing the potentials in the Dirac equation as follows%
\begin{eqnarray}
V_{\Sigma }\left( r\right) &=&\frac{\alpha _{\Sigma }}{r^{2}}+\frac{\beta
_{\Sigma }}{r}+\gamma _{\Sigma }r^{2},  \notag \\
&&  \label{forma} \\
U\left( r\right) &=&\frac{\beta _{u}}{r}+\gamma _{u}r^{\delta _{u}},\quad
\delta _{u}=0\;\text{or\ }1\ .  \notag
\end{eqnarray}%
In these last expressions, when $\delta =2$ one must have $\beta _{\Sigma
}=0\,,\delta _{u}=1$ and when $\delta =-1$ one has $\gamma _{\Sigma
}=0\,,\delta _{u}=0$.

\subsubsection{The effective singular harmonic oscillator}

With $(\delta ,\Lambda )=(2,1/2)$ plus the definition $A=M\omega ^{2}/2$,
the identification of the bound-state solutions of Eq. (\ref{Eq}) with those
ones from the generalized Morse potential is done by setting $V_{1}=-\alpha
^{2}r_{0}^{2}\left( \widetilde{\varepsilon }-C\right) /4$ and $V_{2}=\alpha
^{2}r_{0}^{4}M\omega ^{2}/8$, with $\widetilde{\varepsilon }>C$ and $\omega
^{2}>0$, since $V_{1}<0$ and $V_{2}>0$. With $\omega >0$ one can write%
\begin{equation}
\xi =M\omega r^{2}.
\end{equation}%
Furthermore, (\ref{cond}) implies $\widetilde{\varepsilon }>C+\omega \left(
2n+1\right) $. Using (\ref{ENE}) and (\ref{etil}) one can write the complete
solution of the problem as%
\begin{eqnarray}
\widetilde{\varepsilon } &=&C+\omega \left( 2n+1+S\right)  \notag \\
&&  \label{tilde_eps} \\
g_{\kappa }(r) &=&Nr^{1/2+S}e^{-M\omega r^{2}/2}L_{n}^{\left( S\right)
}\left( M\omega r^{2}\right) .  \notag
\end{eqnarray}%
The condition (\ref{cond}) means that
\begin{equation}
n\leq \bigg[\frac{\widetilde{\varepsilon }-C-\omega }{2\omega }\bigg],
\label{condSHO}
\end{equation}%
where $[x]$ stands for the largest integer less or equal to $x$. Since $%
\widetilde{\varepsilon }$ depends quadratically on $\varepsilon $ from eqs.~(%
\ref{treze}) and $\omega $ may depend at most on $\sqrt{\varepsilon }$ (see
eq. (\ref{param_Morse_osc}) below), the condition (\ref{condSHO}) means that
there is no limitation on the value of $n$, because it can be as large as
the energy can, which in turns means that $n$ in (\ref{tilde_eps}) has no
upper bound.

Examples of this class of solutions can be reached by choosing
\begin{equation}
V_{\Sigma }\left( r\right) =\frac{\alpha _{\Sigma }}{r^{2}}+\gamma _{\Sigma
}r^{2},\quad U\left( r\right) =\frac{\beta _{u}}{r}+\gamma _{u}r.  \label{P1}
\end{equation}%
This is the vector-scalar SHO potential plus the tensor potential Cornell
potential \cite{ham2}, which under appropriate conditions can describe
particular cases like the harmonic oscillator plus a tensor linear potential
\cite{harm_osc_prc_2004}, the harmonic oscillator plus a tensor Cornell
potential \cite{akc1}-\cite{zar2}, the SHO plus a tensor linear potential
\cite{Aydigdu_Sever_2010}, the SHO \cite{isotonic_osc}-\cite{lou}, the
tensor Cornell potential \cite{akc} and the Dirac oscillator \cite{kuku}.

The complete identification with the generalized Morse potential is done
with the equalities%
\begin{eqnarray}
M\omega &=&\sqrt{\gamma _{u}^{2}+\gamma _{\Sigma }\left( \varepsilon
+m\right) }  \notag \\
2MB &=&\left( \beta _{u}+\kappa +1/2\right) ^{2}-\left( \kappa +1/2\right)
^{2}+\alpha _{\Sigma }\left( \varepsilon +m\right)  \label{param_Morse_osc}
\\
2MC &=&\gamma _{u}\left( 2\beta _{u}+2\kappa -1\right) ,  \notag
\end{eqnarray}%
which lead, in general, to an irrational equation in $\varepsilon $:%
\begin{align}
\left( \varepsilon +m\right)&\left( \varepsilon -m\right) -\gamma _{u}\left(
2\beta _{u}+2\kappa -1\right) =2\left( 2n+1+S\right) \sqrt{\gamma
_{u}^{2}+\gamma _{\Sigma }\left( \varepsilon +m\right) }  \notag \\
&=2\left( 2n+1+\sqrt{\left( \beta _{u}+\kappa +1/2\right) ^{2}+\alpha
_{\Sigma }\left( \varepsilon +m\right) }\right) \sqrt{\gamma _{u}^{2}+\gamma
_{\Sigma }\left( \varepsilon +m\right) } .  \label{Na}
\end{align}
We note that if $\alpha_\Sigma>0$, $\gamma_\Sigma>0$ and $\beta_u=0$, one
gets a harmonic oscillator type energy spectrum for positive energy states
with $\varepsilon>m$, but there also states with negative energy, although
there would a minimum value for that energy, because one must have $%
\left(\kappa+1/2\right) ^{2}+\alpha_{\Sigma}\left( \varepsilon +m\right)\geq
0$. If in addition $\alpha_\Sigma=0$, one has the (positive energy)
generalized relativistic harmonic oscillator with $\gamma_{\Sigma
}=1/2\,m\omega_1^2$, $\gamma_u=m\omega_2$ where $\omega_1$ and $\omega_2$
are the frequencies defined in \cite{harm_osc_prc_2004}.

Squaring Eq. (\ref{Na}) successively results into a nonequivalent algebraic
equation of degree $8$. Solutions of this algebraic equation that are not
solutions of the original equation can be removed by backward substitution.
A quartic algebraic equation is obtained when $\alpha _{\Sigma }=0$. For $%
\alpha _{\Sigma }=\gamma _{u}=0$ one obtains a cubic algebraic equation.
However, (\ref{Na}) can be written as a quadratic algebraic equation
rendering two branches of solutions symmetrical about $\varepsilon =0$ in
the case of a pure tensor Cornell potential ($\gamma _{u}\neq 0$):%
\begin{eqnarray}
\varepsilon &=&\pm \sqrt{m^{2}+\gamma _{u}\left( 2\beta _{u}+2\kappa
-1\right) +2|\gamma _{u}|\left( 2n+1+S\right) }  \notag \\
&=&\pm \sqrt{m^{2}+\gamma _{u}\left( 2\beta _{u}+2\kappa -1\right) +2|\gamma
_{u}|\left( 2n+1+|\beta _{u}+\kappa +1/2|\right) }.
\end{eqnarray}

\subsubsection{The effective singular Coulomb potential}

To get the bound states of eq. (\ref{Eq}) from those of the generalized
Morse potential equation eq. (\ref{sch2}) with the pair $(\delta ,\Lambda
)=(-1,1)$ one must choose $V_{1}=\alpha ^{2}r_{0}A$ and $V_{2}=\alpha
^{2}r_{0}^{2}\left( C-\widetilde{\varepsilon }\right) $, with $A<0$ and $%
\widetilde{\varepsilon }<C$. Now,
\begin{equation}
\xi =2\sqrt{2M\left( C-\widetilde{\varepsilon }\right) }\,r
\end{equation}%
and (\ref{cond}) implies $\widetilde{\varepsilon }>C-MA^{2}/[2\left(
n+1/2\right) ^{2}]$. Using (\ref{ENE}) and (\ref{etil}) one can write%
\begin{eqnarray}
\widetilde{\varepsilon } &=&C-\frac{MA^{2}}{2\zeta ^{2}}  \notag \\
&& \\
g_{\kappa }\left( r\right) &=&Nr^{1/2+S}e^{-M|A|r/\zeta }L_{n}^{\left(
2S\right) }\left( \frac{2M|A|r}{\zeta }\right) .  \notag
\end{eqnarray}%
with%
\begin{equation}
\label{zeta}
\zeta =n+1/2+S=n+1/2+\sqrt{\left( \kappa +1/2\right) ^{2}+2MB}.
\end{equation}

Again for this class of effective potentials, condition (\ref{cond}) implies
that there is no upper bound for $n$. This class of solutions can be
obtained by choosing
\begin{equation}
V_{\Sigma }\left( r\right) =\frac{\alpha _{\Sigma }}{r^{2}}+\frac{\beta
_{\Sigma }}{r},\quad U\left( r\right) =\frac{\beta _{u}}{r}+\gamma _{u},
\end{equation}%
which is the vector-scalar SCP plus a shifted Coulomb tensor potential.
There results%
\begin{eqnarray}
2MA &=&\beta _{\Sigma }\left( \varepsilon +m\right) +2\gamma _{u}\left(
\beta _{u}+\kappa \right)  \notag \\
2MB &=&\left( \beta _{u}+\kappa +1/2\right) ^{2}-\left( \kappa +1/2\right)
^{2}+\alpha _{\Sigma }\left( \varepsilon +m\right) \\
2MC &=&\gamma _{u}^{2}.  \notag
\end{eqnarray}%
Subject to appropriate constraints, one finds the irrational equation in $\varepsilon $%
\begin{equation}
\left( \varepsilon +m\right) \left( \varepsilon -m\right) =\gamma _{u}^{2}-
\left[ \frac{2\gamma _{u}\left( \beta _{u}+\kappa \right) +\beta _{\Sigma
}\left( \varepsilon +m\right) }{2\left( n+1/2+\sqrt{\left( \beta _{u}+\kappa
+1/2\right) ^{2}+\alpha _{\Sigma }\left( \varepsilon +m\right) }\right) }%
\right] ^{2},
\end{equation}%
One example of solutions for these type of radial potentials in the Dirac
equation is the Coulomb potential plus a tensor Coulomb potential \cite{ham4}%
, and the SCP plus a tensor Coulomb potential \cite{ham3}-\cite{eshghia}.

The very special case $\alpha_{\Sigma }=\gamma _{u}=0$, necessarily with $%
\beta _{\Sigma }<0$, holds a spectrum given by%
\begin{equation}
\varepsilon =m\frac{1-\left[ \beta _{\Sigma }/\left( 2\zeta \right) \right]
^{2}}{1+\left[ \beta _{\Sigma }/\left( 2\zeta \right) \right] ^{2}},
\end{equation}%
with $\zeta =n+1/2+|\beta _{u}+\kappa +1/2|$. It is interesting that the
dependence on the tensor potential parameter $\beta _{u}$ is done only
through $\zeta $, which contains the quantity $2MB$. Therefore, the spectrum
is formally similar to the solution of pure ($\beta _{u}=0$) Coulomb scalar
and vector potentials in spin symmetry conditions \cite{coulomb_pra}. It
amounts to have an effective value of $\kappa $, given by $\bar{\kappa}%
=\kappa +\beta _{u}$.

It is also interesting to see that the special case $\alpha _{\Sigma }=\beta
_{\Sigma }=\beta _{u}=0$ gives a spectrum for either spin aligned or spin
antialigned, depending on the sign of $\gamma _{u}$.

\subsubsection{Summary of results}

In the following table we summarize the conditions for the potential parameters which allow for analytical solutions.

\begin{table}[th]
\renewcommand{\arraystretch}{1.5}
\begin{center}
$%
\begin{array}{|c|c|c|}
\hline
\delta_u=1 & \delta_u=0 \\
\hline
\beta_\Sigma=0&\gamma_\Sigma=0\\
M^2\omega^2=\gamma _{u}^2+\gamma_\Sigma(\varepsilon+m)>0& 2MA=\beta_\Sigma(\varepsilon+m)+2\gamma_u(\beta_u+\kappa)<0  \\
S=(\beta_u+\kappa+1/2)^2+\alpha_\Sigma(\varepsilon+m)>0 &S=(\beta_u+\kappa+1/2)^2+\alpha_\Sigma(\varepsilon+m)>0\\
g_{\kappa }(r)=Nr^{1/2+S}e^{-M\omega r^{2}/2}L_{n}^{\left( S\right)
}\left( M\omega r^{2}\right)&g_{\kappa }\left( r\right)=Nr^{1/2+S}e^{-M|A|r/\zeta }L_{n}^{\left(
2S\right) }\left(\frac{2M|A|r}{\zeta}\right)\\
\hline
\end{array}%
$%
\end{center}
\caption{General conditions for the potential parameters of eq. (\ref{forma}) in order to have
analytical solutions and radial functions $g_{\kappa }(r)$ for $\delta_u=1$ (harmonic oscillator type potentials)
and for $\delta_u=0$ (Coulomb type potentials). $\zeta$ is given by (\ref{zeta}).}
\end{table}

\subsection{Isolated solutions for $V_{\Delta }=0$ ($\protect\varepsilon =-m$%
)}

We shall now deal with possible solutions for potentials expressed by (\ref%
{forma}) that can not be expressed by means of the Sturm-Liouville problem.
For $V_{\Delta }=0$ and $\varepsilon =-m$, one can write
\begin{eqnarray}
\frac{dg_{\kappa }\left( r\right) }{dr}+\left[ \frac{\kappa }{r}+U\left(
r\right) \right] g_{\kappa }\left( r\right) &=&0  \notag \\
&&  \label{p5} \\
\frac{df_{\kappa }\left( r\right) }{dr}-\left[ \frac{\kappa }{r}+U\left(
r\right) \right] f_{\kappa }\left( r\right) &=&\left[ 2m+V_{\Sigma }\left(
r\right) \right] g_{\kappa }\left( r\right) ,  \notag
\end{eqnarray}%
which arise from (\ref{P5}). Hence,%
\begin{equation}
g_{\kappa }(r)=N_{g}e^{-v\left( r\right) },  \label{gg}
\end{equation}%
with%
\begin{equation}
v\left( r\right) =\int^{r}dy\left[ \frac{\kappa }{y}+U\left( y\right) \right]
.  \label{vv}
\end{equation}%
There is no need to use a lower limit on the integral in (\ref{vv}) because
the resulting constant of integration can be lumped in the constant $N_{g}$.
On the other hand, the nonhomogeneous differential equation for $f_{\kappa }$
yields the general solution
\begin{equation}
f_{\kappa }\left( r\right) =\left[ N_{f}+N_{g}I\left( r\right) \right]
e^{+v\left( r\right) },
\end{equation}%
where $N_{f}$ is a constant associated to the homogeneous equation for $%
f_{\kappa }$, and%
\begin{equation}
I\left( r\right) =\int^{r}dy\left[ 2m+V_{\Sigma }\left( y\right) \right]
e^{-2v\left( y\right) }.
\end{equation}%
It is also worthwhile to note that this sort of isolated solution can not
describe scattering states.

One finds%
\begin{equation}
v\left( r\right) =\ln r^{\left( \beta _{u}+\kappa \right) }+\frac{\gamma
_{u}}{\delta _{u}+1}r^{\delta _{u}+1}.
\end{equation}%
Because $g_{\kappa }$ and $f_{\kappa }$ are square-integrable functions, $%
N_{f}=0$ for $\gamma _{u}>0$, and $N_{g}=0$ for $\gamma _{u}<0$. Hence,%
\begin{eqnarray}
g_{\kappa }\left( r\right) &=&N_{g}r^{-\left( \beta _{u}+\kappa \right)
}\exp \left( -\frac{|\gamma _{u}|}{\delta _{u}+1}r^{\delta _{u}+1}\right)
\notag \\
&& \\
f_{\kappa }\left( r\right) &=&N_{g}I\left( r\right) r^{+\left( \beta
_{u}+\kappa \right) }\exp \left( +\frac{|\gamma _{u}|}{\delta _{u}+1}%
r^{\delta _{u}+1}\right) .  \notag
\end{eqnarray}%
for $\gamma _{u}>0$, and
\begin{eqnarray}
g_{\kappa }(r) &=&0  \notag \\
&& \\
f_{\kappa }\left( r\right) &=&N_{f}r^{+\left( \beta _{u}+\kappa \right)
}\exp \left( -\frac{|\gamma _{u}|}{\delta _{u}+1}r^{\delta _{u}+1}\right) ,
\notag
\end{eqnarray}%
for $\gamma _{u}<0$.

When $\gamma _{u}>0$, square integrability of $f_{\kappa }$ demands a good
behaviour for $I(r)$ at infinity. Calculation shows that

\begin{eqnarray}
&&\left( \delta _{u}+1\right) \left( \frac{2|\gamma _{u}|}{\delta _{u}+1}
\right) ^{-2\left( \beta _{u}+\kappa \right) /\left( \delta _{u}+1\right)
}I\left( r\right)  \notag \\
&&  \notag \\
&=&2m\left( \frac{2|\gamma _{u}|}{\delta _{u}+1}\right) ^{-1/\left( \delta
_{u}+1\right) }\Gamma\left( \frac{1-2\left( \beta _{u}+\kappa
\right) }{\delta _{u}+1},\frac{2|\gamma _{u}|}{\delta _{u}+1}r^{\delta
_{u}+1}\right)  \notag \\
&&  \notag \\
&&+\alpha _{\Sigma }\left( \frac{2|\gamma _{u}|}{\delta _{u}+1}\right)
^{+1/\left( \delta _{u}+1\right) }\Gamma\left( -\frac{1+2\left(
\beta _{u}+\kappa \right) }{\delta _{u}+1},\frac{2|\gamma _{u}|}{\delta
_{u}+1} r^{\delta _{u}+1}\right)  \notag \\
&&  \notag \\
&&+\beta _{\Sigma }\Gamma\left( -\frac{2\left( \beta _{u}+\kappa
\right) }{\delta _{u}+1},\frac{2|\gamma _{u}|}{\delta _{u}+1}r^{\delta
_{u}+1}\right)  \notag \\
&&  \notag \\
&&+\gamma _{\Sigma }\left( \frac{2|\gamma _{u}|}{\delta _{u}+1}\right)
^{-3/\left( \delta _{u}+1\right) }\Gamma\left( \frac{3-2\left(
\beta _{u}+\kappa \right) }{\delta _{u}+1},\frac{2|\gamma _{u}|}{\delta
_{u}+1} r^{\delta _{u}+1}\right)
\end{eqnarray}%
where $\Gamma\left( a,z\right) $ is the incomplete gamma function
\cite{abramowitz}%
\begin{equation}
\Gamma\left( a,z\right) =\int\nolimits_{0}^{z}dt\,e^{-t}t^{a-1}.
\end{equation}%
Due to the behaviour of the integrand near the origin, this integral
diverges if $\text{Re }a$ is not positive. Furthermore, as $z$ increases $%
\Gamma\left( a,z\right) $ approaches the limiting value $\Gamma
\left( a\right) $ when $\text{Re }a>0$. Therefore, $I(r)$ diverges if the
first argument of the incomplete gamma function of at least one of the terms
of $I(r)$ is not positive, and it tends to a constant as $r$ tends to
infinity if the first argument of the incomplete gamma function of all the
terms of $I(r)$ is positive. For these reasons, $f_{\kappa }$ is not a
square-integrable function. An exception, though, occurs when $m=\alpha
_{\Sigma }=\beta _{\Sigma }=\gamma _{\Sigma }=0$ just for the reason that $%
f_{\kappa }$ vanishes identically. Therefore,%
\begin{eqnarray}
g_{\kappa }\left( r\right) &=&N_{g}r^{-\left( \beta _{u}+\kappa \right)
}\exp \left( -\frac{|\gamma _{u}|}{\delta _{u}+1}r^{\delta _{u}+1}\right)
\notag \\
&& \\
f_{\kappa }\left( r\right) &=&0,  \notag
\end{eqnarray}%
only for $\gamma _{u}>0$ and $m=V_{\Sigma }=0$.

In addition, a good behaviour of $g_{\kappa }$ and $f_{\kappa }$ near the
origin, in the sense of normalization, forces one to the choice $\beta
_{u}+\kappa \lessgtr \pm 1/2$ for $\gamma _{u}\gtrless 0$.

\section{Concluding remarks}

Based on Ref. \cite{new}, we have described a straightforward and efficient
procedure for finding a large class of solutions of the Dirac equation in
3+1 dimensions with radial scalar $V_{s}$ vector $V_{v}$ and tensor $U$
radial potentials, when $V_{s}=\pm V_{v}$, some of which have never been
obtained before. Their wave functions are all expressed in terms of
generalized Laguerre polynomials and their energy eigenvalues obey
analytical equations, either polynomial or irrational which can be cast as
polynomial. These include harmonic oscillator-type and Coulomb-type
potentials and their extensions. Although the solutions for those systems
could be found by standard methods, this procedure, based on the mapping
from the one-dimensional generalized Morse potential via a Langer
transformation to the radial Dirac equation in $3+1$ dimensions, provides an
easier and powerful way to find the solutions of a class of
potentials which otherwise one might not know that would have analytical
solutions in the first place. We were able to reproduce well-known
particular cases of relativistic harmonic oscillator and Coulomb spin-1/2
systems, when the scalar and vector potentials have the same magnitude, but
there are a wealth of other particular cases with physical interest that are
left for further study, one of them being solutions with Coulomb-type potentials with
tensor components.

\section*{Acknowledgement}

This work was supported in part by means of funds provided by CAPES and CNPq
(grants 455719/2014-4, 304105/2014-7 and 304743/2015-1). PA would like to
thank the Universidade Estadual Paulista, Guaratinguet\'a Campus, for
supporting his stays in its Physics and Chemistry Department and CFisUC for travel support.

\end{document}